\documentclass[twocolumn,apj,iop]{openjournal}

\DeclareRobustCommand{\VAN}[3]{#2}
\let\VANthebibliography\thebibliography
\def\thebibliography{\DeclareRobustCommand{\VAN}[3]{##3}\VANthebibliography}

\usepackage{graphicx}	

\usepackage{orcidlink}

\newcommand{\newtable}{\relax}

\begin{document}

\title{Optimal Summary Statistics for X-ray Polarization}
\shorttitle{Optimal Statistics}
\author{Jeremy Heyl\orcidlink{0000-0001-9739-367X}$^1$}
\author{Denis Gonz{\'a}lez-Caniulef\orcidlink{0000-0001-5848-0180}$^2$}
\author{Ilaria Caiazzo\orcidlink{0000-0002-4770-5388}$^3$}
\shortauthors{Heyl et al.}

\affiliation{$^1$University of British Columbia, Vancouver, BC, Canada}
\affiliation{$^{2}$Institut de Recherche en Astrophysique et Plan\'etologie, UPS-OMP, CNRS, CNES, 9 avenue du Colonel Roche, BP 44346 31028, Toulouse CEDEX 4, France}
\affiliation{$^3$California Institute of Technology,
Pasadena, CA, USA}

\begin{abstract}
We develop two new highly efficient estimators to measure the polarization (Stokes parameters) in experiments that constrain the position angle of individual photons such as scattering and gas-pixel-detector polarimeters, and analyse in detail a previously proposed estimator. All three of these estimators are at least fifty percent more efficient on typical datasets than the standard estimator used in  the field.  We present analytic estimates of the variance of these estimators and numerical experiments to verify these estimates.  Two of the three estimators can be calculated quickly and directly through summations over the measurements of individual photons. 
\end{abstract}

\keywords{
methods: data analysis -- methods: statistical -- techniques: polarimetric -- X-rays: general}

\maketitle


\section{Introduction}

With the advent of X-ray polarimeters such as the Imaging X-Ray Polarimetry Explorer \citep[IXPE][]{2022JATIS...8b6002W} and XL-Calibur \citep{2021APh...12602529A}, techniques to estimate the polarized flux from astrophysical sources using measurements of the scattering or photo-electric emission induced by individual photons are in high demand.  \citet{Kislat2015} developed the standard estimator in the field which has several conceptual and technical advantages.  The Kislat estimator assigns Stokes parameters to individual photons, so the total Stokes parameters derived from a particular measurement is simply the sum over those of the individual photons.  Although this estimator is unbiased, it is far from optimal, especially when considering an instrument whose sensitivity to polarization varies from photon to photon. This is the case for current X-ray polarimeters, which provide a sinusoidal polarization signal whose amplitude is proportional to a modulation factor ($\mu$) that depends strongly on energy \citep{2021ApJ...907...82M}. The efficiency of an estimator is inversely proportional to the number of photons required to achieve an expected signal-to-noise ratio; for example, the Kislat estimator would require a fifty percent longer observation (or a fifty percent larger telescope) to obtain the same information as an estimator fifty percent more efficient. 

In this paper, we revisit the problem of the optimal summary statistics to compute the Stokes parameters from X-ray polarimetric observations. The paper is organized as follows. In Section~\ref{sec:deriv} we start with the theoretically most efficient estimator, the maximum likelihood estimator \citep{gonzalezcaniulef_unbinned} or MLE, and derive two linear estimators from it.  Although the MLE yields the smallest variance (or highest signal-to-noise ratio), it must be calculated iteratively through an optimization process which involves calculating the likelihood of each of many hundreds of thousands or millions of photons repeatedly (perhaps hundreds of times), and so can be very time consuming.  On the other hand, the two linear estimators can be calculated through sums over the properties of individual photons.  While the Kislat estimator requires two summations, one for each of the two Stokes parameters for linear polarization, the efficient estimators that we present here require five or three summations over the photons, so the added computational effort to calculate these linear estimators is negligible compared to a full MLE analysis.  In Section~\ref{sec:results} we perform numerical experiments to assess the efficiency of each estimator and verify the analytic calculations.  Conclusions are presented in Section~\ref{sec:conclus}.

\section{Derivations}
\label{sec:deriv}
We can build the likelihood function focusing only on the polarization signal, which is a function of photon angle, energy and time, or by including the spectrum, which is just a function of energy and time, in the model as well \citep[see also][]{Kislat2015,2021ApJ...907...82M,2021AJ....162..134M}. Let $p_m(E,t)$ be polarization degree of the model, $\mu(E)$ be the modulation factor of the instrument and $\psi_m(E,t)$ be the polarization angle of the model; the first component of the likelihood function consists of the probability density of the photon position angle which can therefore be written as:
\begin{equation}
    f(\psi) = \frac{1}{2\pi} \left [  1 +  \mu p_m \left( 2 \cos^2 (\psi-\psi_m)-1 \right)  \right]
\end{equation}
where $\psi$ is the position angle (or photo-electron scattering angle) of a particular photon and the energy and time variables are suppressed.  This expression results from the definition of the modulation factor, the differential scattering cross-section, and the normalization $\int f(\psi) d \psi=1$, which is constant with respect to the expected degree of polarization.  In principle, the detector could introduce some uncertainty in the measured value of the angle $\psi$; if the detector is unbiased, the only effect of this is to reduce the value of $\mu$ from the theoretical value \citep{2021AJ....162..134M}.  We will assume that the value of $\mu$ has been measured or calculated including these angular uncertainties.

We can reformulate the probability density for a single photon as as \citep{2021ApJ...907...82M} 
\begin{equation}
f(\psi) = \frac{1}{2\pi} \left  [ 1 + \mu p_m \left (  \cos2\psi \cos2\psi_m + \sin2\psi\sin2\psi_m \right ) \right ].
\end{equation}
\citep{BESSET1979515} use an analogous probability density for spin-half particles and derive analogous expressions (with $2\psi$ replaced by $\psi$). If we define $q_m=p_m\cos2\psi_m$ and $u_m=p_m\sin2\psi_m$ for the model, and $Q_\gamma= \cos2\psi$ and $U_\gamma=\sin2\psi$ for the photon, we have
\begin{equation}
f_\gamma =  \frac{1}{2\pi} \left  [1 + \mu_\gamma \left ( Q_\gamma q_m + U_\gamma u_m \right ) \right ]  =  \frac{1}{2\pi} \left  [1 + \mu_\gamma {\bf S}_\gamma \cdot {\bf s}_m  \right ] 
\end{equation}
where we have placed the normalized Stokes parameters ($q_m$ and $u_m$ and similarly for the photon) into a two-component Stokes vector (${\bf s}_m$) for notational convenience. In general we will use lowercase letters to signify normalized Stokes parameters; that is, $q=Q/I$ etc.  For IXPE, the total logarithmic likelihood of the data given the model is
\begin{equation}
\log L = \sum_\gamma \log f_\gamma - N_\textrm{\scriptsize pred}\,,
\end{equation}
where  $N_\textrm{\scriptsize pred}$ is the number of photons predicted by the model (it does not depend on the assumed model polarization).  We can maximize the likelihood to determine the most-likely values of $q_m$ and $u_m$ to account for the data which we will denote as ${\bf s}_{m,\textrm{\scriptsize MLE}}$. Although this process is necessarily iterative, it has the advantage of yielding the minimum-variance unbiased estimator for the polarization as the MLE achieves the Cram\'er-Rao lower bound on the variance \citep{cramer,rao}.  Two useful approximations are the expected values
\begin{equation}
\langle \log(1+\mu_\gamma q_m \cos2\psi) \rangle \approx \frac{1}{4}(\mu_\gamma q_m)^2 + \frac{1}{16} (\mu_\gamma q_m)^4 + \cdots
\end{equation}
and its variance which is given by
\begin{equation}
\textrm{Var} \left [ \log(1+\mu_\gamma q_m \cos2\psi) \right ]\approx \frac{1}{2} (\mu_\gamma q_m)^2 - \frac{3}{32}(\mu_\gamma q_m)^4 + \cdots
\end{equation}
to determine how well converged the iterative process is.  As the likelihood is calculated as the sum over a large number of terms (one for each photon) and each one is an independent and identically distributed random variable drawn from a distribution with a finite mean and variance, the distribution of the sum approaches the normal distribution by the central limit theorem. Therefore, using the expected value for the likelihood and a odds ratio of 99, we can estimate the minimum detectable polarization at 99\% confidence to be
\begin{equation}
\textrm{MDP}_{99} = 2\sqrt{\frac{\log(99)}{N\langle\mu^2\rangle}} \approx \frac{4.29}{(N \langle\mu^2\rangle )^{1/2}}.
\end{equation}
where $N$ is the number of events, in agreement with \citet{2021ApJ...907...82M}.

We would like to find a direct estimator of the polarization that approaches the variance of the MLE and connects with the estimator traditionally used in the field \citep{Kislat2015}.  Our first step is to take the gradient of the likelihood with respect to the model polarization ${\bf S}_m$ and linearize it,
\begin{eqnarray}
\nabla_{{\bf s}_m} \log L &=& \sum_\gamma \frac{\mu_\gamma {\bf S}_\gamma}{1 + \mu_\gamma {\bf S}_\gamma \cdot {\bf s}_m } \\
&\approx& \sum_\gamma \mu_\gamma {\bf S}_\gamma \left [ 1 - \mu_\gamma {\bf S}_\gamma \cdot {\bf s}_m \right ].
\end{eqnarray}
In linearizing the likelihood we have neglected terms of higher order in $\mu_\gamma \bf{s}_m$, so we have introduced a potential bias for highly polarized sources observed with instruments with high modulation factors (the current generation of instruments typically have $\mu<0.5$ so this is not an issue at this point). To maximize the likelihood we have
\begin{equation}
\sum_\gamma \mu_\gamma  {\bf S}_\gamma  \approx \sum_\gamma  \mu_\gamma^2 {\bf S}_\gamma  {\bf S}_\gamma \cdot {\bf s}_m 
\end{equation}
and
\begin{equation}
{\bf s}_{m,1} =\left [  \sum_\gamma  \mu_\gamma^2 {\bf S}_\gamma \otimes  {\bf S}_\gamma \right ]^{-1} \sum_\gamma \mu_\gamma  {\bf S}_\gamma  \label{eq:Sm1}
\end{equation}
where $\otimes$ denotes the Kronecker product. This estimator\footnote{It requires 5 summations over the events that can number more than a million to compute the Stokes.} is 
the same as derived by \citet[][Eq.~59 and~60]{2021ApJ...907...82M} and 
similar to the approximate estimator \citep[][Eq.~61 and~62]{2021ApJ...907...82M} 
\begin{equation}
q_{m,M} =\frac{ \sum_\gamma \mu_\gamma  Q_\gamma  }{\sum_\gamma \mu_\gamma^2  Q_\gamma^2  }\label{eq:qmM}
\end{equation}
and similarly for $u_{m,M}$. If we replace the term in the brackets with its expectation value, we obtain\footnote{It requires 3 summations to compute the Stokes.}
\begin{equation}
{\bf s}_{m,2} = 2 \frac{\sum_\gamma \mu_\gamma  {\bf S}_\gamma  }{  \sum_\gamma  \mu_\gamma^2 }
\end{equation}
which is very similar to the \citet{Kislat2015} result,
\begin{equation}
{\bf s}_{m,K} = 2 \sum_\gamma \frac{{\bf S}_\gamma}{\mu_\gamma}.
\end{equation}
but with a different weighting by modulation factor. The Kislat estimator is implemented in \texttt{ixpeobssim} \citep{2022SoftX..1901194B} and \texttt{xselect} \citep{1996ASPC..101...17A}. We can understand the relationship between the ${\bf s}_{m,K}$ and ${\bf s}_{m,2}$ estimators by grouping the observations according to the values of $\mu_\gamma$ and calculating ${\bf s}_{m,K}$ for each group. For each group the variance in the estimator is
\begin{equation}
\text{Var} ( q_{m,K} ) = \frac{1}{N_\mu} \left ( \frac{2}{\mu^2} - q_{m,K,\mu}^2 \right )
\end{equation}
and similarly for $u_{m,K}$. If we combine the individual estimators $q_{m,K,\mu}$ weighted inversely by the variance, we get
\begin{equation}
q_{m,\textrm{\scriptsize weighted mean}} = \frac{\sum_\mu N_{\mu}\mu^2 q_{m,K,\mu}}{\sum_\mu N_\mu \mu^2} = 2 \frac{\sum_\gamma \mu_\gamma  Q_\gamma  }{  \sum_\gamma  \mu_\gamma^2 } = q_{m,2}
\end{equation}
where we have ignored the second term in the variance.  In a similar fashion we can estimate the variance in the two estimators as
\begin{equation}
\textrm{Var}(q_{m,2})=\frac{1}{N} \left ( \frac{2}{\langle \mu^2\rangle} - q_{m,2}^2 \right )
\end{equation}
and
\begin{equation}
\textrm{Var}(q_{m,K}) = \frac{1}{N} \left ( \left\langle \frac{2}{\mu^2} \right\rangle - q_{m,K}^2 \right ).
\end{equation}
The two preceding equations contain the key result of this work.  Because the root-mean-square of a group of non-identical quantities is greater than the harmonic-root-mean-square, the variance of ${\bf s}_{m,2}$ is necessarily smaller than that of ${\bf s}_{m,K}$ if the modulation factors of the observed photons vary.

We can also estimate the variance for the MLE technique from the second derivative of the log-likelihood
\begin{equation}
\frac{\partial^2 \log L}{\partial {\bf s}_m \partial {\bf s}_m} = \sum_\gamma \frac{\mu_\gamma^2 {\bf S}_\gamma \otimes  {\bf S}_\gamma}{\left (1 + \mu_\gamma {\bf S}_\gamma \cdot {\bf s}_m \right )^2}.
\end{equation}
With no loss of generality, let us take $\mu q_m=\sin\alpha$ and $u_m=0$ to calculate the following expectation values 
\begin{equation}
\left \langle \frac{\partial^2 \log L}{\partial q_m \partial q_m} \right \rangle = N \left \langle \frac{ \mu^2}{\cos\alpha + \cos^2\alpha} \right \rangle
\end{equation}
and
\begin{equation}
\left \langle \frac{\partial^2 \log L}{\partial u_m \partial u_m} \right  \rangle = N \left \langle \frac{\mu^2}{1 + \cos\alpha} \right \rangle.
\end{equation}
If we now take the general situation of $q_m\neq 0$ and $u_m\neq 0$ with $\sin^2\alpha = \mu^2 (q_m^2+u_m^2)$ the variances of the two Stokes parameters are
\begin{eqnarray}
\textrm{Var}(q_{m,\textrm{\scriptsize MLE}})&=&\frac{1}{N} \Biggr [ \left \langle \frac{\mu^2}{\cos\alpha+\cos^2\alpha} \right \rangle^{-1} \frac{q_m^2}{q_m^2+u_m^2} + \nonumber \\
& & \left \langle \frac{\mu^2}{1+\cos\alpha} \right \rangle^{-1} \frac{u_m^2}{q_m^2+u_m^2}\Biggr ]
\end{eqnarray}
and
\begin{eqnarray}
\textrm{Var}(u_{m,\textrm{\scriptsize MLE}})&=&\frac{1}{N} \Biggr [ \left \langle \frac{\mu^2}{\cos\alpha+\cos^2\alpha} \right \rangle^{-1} \frac{u_m^2}{q_m^2+u_m^2} +  \nonumber \\
& & \left \langle \frac{\mu^2}{1+\cos\alpha} \right \rangle^{-1} \frac{q_m^2}{q_m^2+u_m^2}\Biggr ]
\end{eqnarray}

The variance of the remaining estimator ${\bf s}_{m,1}$ is obtained by linearising the results for the MLE to yield
\begin{equation}
\textrm{Var}(q_{m,1})=\frac{1}{N \langle \mu^2\rangle} \left [ 2 -  \left  \langle \frac{1}{\mu^2} \right \rangle^{-1} \left ( \frac{3}{2} q_{m,1}^2 +\frac{1}{2} u_{m,1}^2 \right )  \right ]
\end{equation}
and
\begin{equation}
\textrm{Var}(u_{m,1})=\frac{1}{N \langle \mu^2\rangle} \left [ 2 -  \left  \langle \frac{1}{\mu^2} \right \rangle^{-1} \left ( \frac{3}{2} u_{m,1}^2 +\frac{1}{2} q_{m,1}^2 \right )  \right ]
\end{equation}
In general the covariance between $Q$ and $U$ for all of the estimators is given by
\begin{equation}
\textrm{Cov}(qu) = -\frac{qu}{N}
\end{equation}

\section{Results}
\label{sec:results}

We perform a series of numerical experiments with a constant number of photons with either a constant modulation factor or one that varies from photon to photon from 0.2 to 0.5 uniformly to determine the properties of the MLE and the direct estimators: ${\bf s}_{m,1},{\bf s}_{m,2}$ and ${\bf s}_{m,K}$.  For the first experiment we take the values of $q=u=0.5$ and $\mu=1$ and ten thousand random realisations each of one thousand events, yielding error estimates of 0.0382, 0.0387, 0.0418 and 0.0418 for the ${\bf s}_{m,\textrm{\scriptsize MLE}}$, ${\bf s}_{m,1}$,${\bf s}_{m,2}$ and the Kislat estimator ${\bf s}_{m,K}$ from the formulae above which agree with the results from the numerical simulations shown in Tab.~\ref{tab:mean}.

In general the modulation factors of individual photons differ \citep[as pointed out by][]{2021ApJ...907...82M}, so we repeat the experiment with modulation factors varying from 0.2 to 0.5 drawn from a uniform distribution.  Although the mean value of $\mu$ is 0.35, the mean value of $\mu^2$ is $(0.361)^2$, and the mean value of $\mu^{-2}$ is $(0.316)^{-2}$, yielding error estimates for the ${\bf s}_{m,1}$, ${\bf s}_{m,2}$ and ${\bf s}_{m,K}$ estimators of 0.122, 0.123 and 0.141 respectively. The minimum variance estimator, ${\bf s}_{m,\textrm{\scriptsize MLE}}$, yields 0.122.  All are in agreement with the simulated results (Tab.~\ref{tab:mean}).  In the third experiment we take the model polarization to vanish and the distribution of modulation factors, yielding uncertainties of 0.142 for the Kislat estimator ${\bf s}_{m,K}$ and 0.124 (see Tab.~\ref{tab:mean}) for the others with the corresponding MDP values of 0.431 and 0.376 (in agreement with the simulation results in Tab.~\ref{tab:mdp}).  The expected values for the covariance in the first two simulations are $-2.5 \times 10^{-4}$ and zero in the final simulation.  The convergence of the simulations (Tab.~\ref{tab:covar}) to the theoretical expectations is poorer than in the case of the standard deviations, but close in magnitude and correct in sign.



\ifx\newtable\undefined
\begin{table*}
  \caption{Results: Mean Values, Standard Deviations.  The MLE technique returns an error estimate in terms of the interval where $\Delta \log L=1$ which is reported in parenthesis.}
  {
  \begin{tabular}{cccccccc}
    \hline
    $q_{m,\textrm{\scriptsize MLE}}$ & $U_{m,\textrm{\scriptsize MLE}}$ & $q_{m,1}$ & $U_{m,1}$ &  $q_{m,2}$ & $U_{m,2}$ &  $q_{m,K}$ & $U_{m,K}$ \\ \hline
    $ \phantom{-}0.500\pm 0.038(0.038)$ & $ 0.500\pm 0.038(0.038)$ & $ \phantom{-}0.500\pm 0.039$ & $ 0.500\pm 0.039$ & $ \phantom{-}0.500\pm 0.042$ & $ 0.500\pm 0.042$ & $\phantom{-}0.500\pm 0.042$ & $ 0.500\pm 0.042$ \\
    $ \phantom{-}0.500\pm 0.122(0.182)$ & $ 0.501\pm 0.122(0.182)$ & $ \phantom{-}0.500\pm 0.122$ & $ 0.501\pm 0.121$ & $ \phantom{-}0.500\pm 0.123$ & $ 0.501\pm 0.123$ & $\phantom{-}0.499\pm 0.141$ & $ 0.500\pm 0.140$ \\
               $-0.001\pm 0.124(0.718)$ & $ 0.000\pm 0.124(0.718)$ & $           -0.001\pm 0.124$ & $ 0.000\pm 0.124$ & $           -0.001\pm 0.124$ & $ 0.000\pm 0.124$ & $-0.000\pm 0.142$ & $ 0.000\pm 0.142$ 
  \end{tabular}
  }
\end{table*}

\begin{table}
    \caption{The measured minimum detectable polarization MDP$_{99}$ for the estimators discussed in the text as determined through the simulations.}
  \begin{tabular}{cccc}
    \hline
    $\textrm{MDP}_{99,\textrm{\scriptsize MLE}}$ & $\textrm{MDP}_{99,1}$ & $\textrm{MDP}_{99,2}$ & $\textrm{MDP}_{99,K}$ \\
    \hline
    0.37486 & 0.37485 & 0.37480 & 0.43172
  \end{tabular}
\end{table}

\begin{table}
  \caption{Results: Covariances between $Q$ and $U$ estimators. The MLE technique returns a covariance estimate in terms of the inverse Hessian of the likelihood function which is reported in parenthesis.  These results are reported in units of $10^{-4}$.}
  \begin{tabular}{cccc}
    \hline
    $\textrm{Cov}(QU)_{m,\textrm{\scriptsize MLE}}$ & $\textrm{Cov}(QU)_{m,1}$ & $\textrm{Cov}(QU)_{m,2}$ & $\textrm{Cov}(QU)_{m,K}$ \\ \hline
    $-2.4585(\phantom{6}-2.500) $ & $-2.4643$ & $          -2.4303$ & $-2.4303$ \\
    $-3.3641(\phantom{-}62.589) $ & $-3.7803$ & $         -3.2506$ & $-2.9467$ \\
    $-0.1347(\phantom{-}96.708) $ & $-0.1354$ & $\phantom{-}0.0116$ & $-0.0169$                                                   
  \end{tabular}
\end{table}
\begin{table*}
  \caption{Results: Mean Values, Standard Deviations, Covariances from Simulated Distributions.
    The results for the $u$ Stokes parameter are identical within the statistical uncertainties. The values of the covariance are reported in units of $10^{-4}$.}
  {
  \begin{tabular}{cccccccc}
    \hline
    $q_{m,\textrm{\scriptsize MLE}}$        & $q_{m,1}$                      &  $q_{m,2}$                   &  $q_{m,K}$                    &      $\textrm{Cov}(QU)_{m,\textrm{\scriptsize MLE}}$ & $\textrm{Cov}(QU)_{m,1}$ & $\textrm{Cov}(QU)_{m,2}$ & $\textrm{Cov}(QU)_{m,K}$ \\ \hline
    $ \phantom{-}0.500\pm 0.038$ & $ \phantom{-}0.500\pm 0.039$  & $ \phantom{-}0.500\pm 0.042$ &  $\phantom{-}0.500\pm 0.042$ &            $-2.4585$ & $-2.4643$ & $          -2.4303$ & $-2.4303$ \\                                                            
    $ \phantom{-}0.500\pm 0.122$ & $ \phantom{-}0.500\pm 0.122$  & $ \phantom{-}0.500\pm 0.123$ &  $\phantom{-}0.499\pm 0.141$ &            $-3.3641$ & $-3.7803$ & $         -3.2506$ & $-2.9467$ \\                                                             
               $-0.001\pm 0.124$ & $           -0.001\pm 0.124$  & $           -0.001\pm 0.124$ &  $-0.000\pm 0.142$           &            $-0.1347$ & $-0.1354$ & $\phantom{-}0.0116$ & $-0.0169$                                                               
  \end{tabular} 
  }
\end{table*}
\else

\begin{table}
  \caption{Results: Mean Values and Standard Deviations from Simulated Distributions for Three Different Experiments (see text for details).
    All of the estimators are apparently unbiased, and 
    the results for the $U$ Stokes parameter are identical within the statistical uncertainties. }\label{tab:mean}
  \centering
  \begin{tabular}{cccc}
    \hline
    $q_{m,\textrm{\scriptsize MLE}}$        & $q_{m,1}$                      &  $q_{m,2}$                   &  $q_{m,K}$                  \\ \hline
    $ 0.50\pm 0.04$ & $ 0.50\pm 0.04$  & $ 0.50\pm 0.04$ &  $0.50\pm 0.04$        \\                                                            
    $ 0.50\pm 0.12$ & $ 0.50\pm 0.12$  & $ 0.50\pm 0.12$ &  $0.50\pm 0.14$           \\                                                             
    $ 0.00\pm 0.12$ & $ 0.00\pm 0.12$  & $ 0.00\pm 0.12$ &  $0.00\pm 0.14$                                                              
  \end{tabular}   
\end{table}

\begin{table}
    \caption{Results: MDP$_{99}$. The measured minimum detectable polarization MDP$_{99}$ for the estimators discussed in the text as determined through the simulations by determining the 99th percentile over the unpolarized simulations.}\label{tab:mdp}
    \centering
  \begin{tabular}{cccc}
    \hline
    $\textrm{MDP}_{99,\textrm{\scriptsize MLE}}$ & $\textrm{MDP}_{99,1}$ & $\textrm{MDP}_{99,2}$ & $\textrm{MDP}_{99,K}$ \\
    \hline
    0.37486 & 0.37485 & 0.37480 & 0.43172
  \end{tabular}
\end{table}
\begin{table}
  \caption{Results: Covariances from Simulated Distributions.
     The values of the covariance are reported in units of $10^{-4}$.}  \label{tab:covar}
  {
  \begin{tabular}{cccc}
    \hline
      $\textrm{Cov}(qu)_{m,\textrm{\scriptsize MLE}}$ & $\textrm{Cov}(qu)_{m,1}$ & $\textrm{Cov}(qu)_{m,2}$ & $\textrm{Cov}(qu)_{m,K}$ \\ \hline
              $-2.4585$ & $-2.4643$ & $          -2.4303$ & $-2.4303$ \\                                                            
            $-3.3641$ & $-3.7803$ & $         -3.2506$ & $-2.9467$ \\                                                             
           $-0.1347$ & $-0.1354$ & $\phantom{-}0.0116$ & $-0.0169$                                                               
  \end{tabular} 
  }
\end{table}
\fi

Let us now apply these formulae to a real dataset in particular the observations of Centaurus X-3 \citep{2022ApJ...941L..14T} with the Imaging X-Ray Polarimetry Explorer \citep{2022JATIS...8b6002W}. The mean value of the modulation factor for the events from the source between two and eight keV is 0.30, the root-mean-square modulation factor is 0.32 and harmonic-root-mean-squared value is 0.26; consequently, to achieve the same constraints on the polarization using the Kislat estimator that is achieved with the estimators presented here requires a fifty-percent longer observation as $(0.32/0.26)^2\approx 1.5$. 

\section{Conclusions} 
\label{sec:conclus}

We have analysed three  estimators for measuring the Stokes parameters in photon counting instruments such as X-ray polarimeters.  All three are about fifty percent more efficient for typical datasets than the estimators implemented in the standard software packages.   One estimator, the ${\bf s}_{m,\textrm{\scriptsize MLE}}$, is theoretically the most efficient; however, it must be calculated iteratively.  The ${\bf s}_{m,1}$ \citep[first derived by][]{2021ApJ...907...82M} is derived by linearizing the MLE and its performance except for very highly polarized sources with highly efficient polarimeters (the product of modulation factor and polarization degree greater than ninety percent) is nearly equivalent to the MLE.  In the regime where the MLE is notably more efficient theoretically than the ${\bf s}_{m,1}$ estimator, the likelihood function becomes very strongly and sharply peaked making the iterations poorly behaved, so in practice even in this regime  the ${\bf s}_{m,1}$ estimator is preferable over the MLE. Furthermore, this estimator saturates the Cram\'er-Rao lower bound on the variance to second order in the polarization degree; therefore, it is the most efficient estimator that can be constructed from the first- and second-order moments of the observed ${\bf S}_\gamma$.

We obtained the second direct estimator ${\bf s}_{m,2}$ by replacing a portion of the expression of the ${\bf s}_{m,1}$ estimator with its expectation value.  Again except for very polarized sources, it is nearly as efficient as ${\bf s}_{m,1}$, but it has the advantage of being straightforward to implement in standard pipelines.  Although the estimators that we have derived here are significantly more efficient than the Kislat estimator  ${\bf s}_{m,K}$ for calculating summary statistics for binned data, the unbinned estimator derived by \citet{gonzalezcaniulef_unbinned} remains the most efficient method for model testing because no binning is performed. 

Although we have not presented calculations of the \citet{2021ApJ...907...82M} approximate estimator, its performance with four summations over photon measurement to derive two Stokes parameters, is in general intermediate between ${\bf s}_{m,1}$ and ${\bf s}_{m,2}$ (indeed almost precisely halfway in between). Additionally an estimator as efficient as ${\bf s}_{m,2}$ can be obtained by calculating the mean Stokes parameters over very narrow energy bins (over which the modulation factor is approximately constant) using the standard Kislat estimator, and taking the average over these bins weighted by the inverse variance to obtain estimates of the Stokes parameters.  

The key point is that the weighting of the photon measurements with modulation factor matters.  The Kislat method yields an uncertainty that is inversely proportional to the harmomic-root-mean-squared modulation factor of the photon measurements, whereas the methods presented here yield uncertainties that are inversely proportional to the root-mean-squared modulation factor. These uncertainties are typically about twenty percent smaller; therefore, these estimators typically yield a fifty percent increase in efficiency,  making a given observation as effective as one fifty percent longer or making a given telescope as effective as one with a fifty percent larger collecting area.

\section*{Acknowledgements}

JH acknowledges support from the Natural Sciences and Engineering Research Council of Canada (NSERC) through a Discovery Grant, the Canadian Space Agency through the co-investigator grant program, and computational resources and services provided the SciServer science platform (www.sciserver.org).
DGC acknowledges support from a CNES fellowship grant.
\section*{Data Availability}

The data used for the Centaurus X-3 observations are available through HEASARC under IXPE Observation ID 01250201.



\bibliographystyle{mnras}
\bibliography{main} 

\end{document}